\documentclass[twocolumn]{pasj00}
\tenpoint

\usepackage{lscape}

\SetRunningHead{T. Miyakawa et al.}{Suzaku observations of MCG-6-30-15}
\title{Spectral Variation of  the Seyfert 1 Galaxy MCG-6-30-15  \\ observed
with Suzaku }
\author{Takehiro \textsc{Miyakawa}\altaffilmark{1,2},
Ken \textsc{Ebisawa}\altaffilmark{1,2},
Yuichi \textsc{Terashima} \altaffilmark{3},
Fuminori \textsc{Tsuchihashi}\altaffilmark{3},  Hajime \textsc{Inoue}\altaffilmark{1}\\
and Piotr \.Zycki\altaffilmark{4}}%
\altaffiltext{1}{The Institute of Space and Astronautical Science (ISAS), Japan
Aerospace Exporlation Agency (JAXA), \\ 3-1-1 Yoshinodai, Sagamihara, Kanagawa 229-8510}
\altaffiltext{2}{Department of Astronomy, Graduate School of Science,
the University of Tokyo,\\
7-3-1 Hongo, Bunkyo-ku, Tokyo 113-0033}
\altaffiltext{3}{
Department of Physics, Ehime University, 2-5 Bunkyocho, Matsuyama, Ehime 790-8577
}
\altaffiltext{4}{
Nicolaus Copernicus Astronomical Center, Bartycka, 18, Warsaw, Poland
}

\email{miyakawa@astro.isas.jaxa.jp}
\Received{$\langle$2009 June 16$\rangle$}
\Accepted{$\langle$2009 $\rangle$}
\Published{$\langle$publication date$\rangle$}
\begin{document}
\maketitle

\begin{abstract}
We have investigated spectral variation
of the Seyfert 1 galaxy  MCG-6-30-15 observed with 
Suzaku in  January 2006 for three separate  periods spreading 
over fourteen days.  We found that the time-averaged continuum energy spectrum between 1 keV
and 40 keV  can be approximated with a spectral model composed of the
direct power-law component, its reflection component,
two  warm absorbers with different ionization states, and neutral absorption.
We have taken two approaches to study its spectral
variation at various timescales:  The first approach is to make intensity-sliced spectra and study
correlation between the intensity and spectral shape.
The second approach is to study spectral changes between the intervals when
the source flux is above (``bright state'') and below 
(``faint state'') the average for  fixed  time-intervals.
In both approaches, we found a clear correlation 
between the intensity in the 6 -- 10 keV band and the spectral
ratio of 0.5 -- 3.0 keV/6.0-- 10 keV.
Such a spectral variation requires change of 
the apparent  slope of the  direct component, 
 whereas the shape and intensity of 
the reflection component being invariable.
The observed apparent spectral change is  explained by 
variation of the ionization degree  of one of the two warm absorbers due to
intrinsic source luminosity  variation.  Current results suggest that 
the warm absorber has a critical role to explain the observed continuum spectral shape and
variation of MCG-6-30-15, which is essential 
to constrain parameters of the putatively  broadened iron line emission feature.

\end{abstract}

\section{Introduction}

The Seyfert 1 galaxy MCG-6-30-15 has been extensively studied 
with   ASCA
(Iwasawa et al. 1996,1999; Matsumoto et al. 2003), BeppoSAX (Guainazzi et
al. 1999), Rossi X-ray Timing Explorer (Lee et al. 1999, Vaughan
$\&$ Edelson 2001), Chandra (Lee et al. 2001, Young et al. 2005, Gibson
et al. 2007), XMM-Newton
(Fabian et al. 2002, Vaughan $\&$ Fabian 2004, Ponti et al. 2004), 
Suzaku (Miniutti et al. 2007), and a combination of
Suzaku, Chandra and XMM-Newton (Miller, Turner, and Reeves 2008). 
ASCA discovered 
a broad and skewed emission line feature around 5--7 keV in the
spectrum of MCG-6-30-15  for the first time   (Tanaka
et al. 1995). 
The iron line is emitted as a part of the reflection 
spectrum from the accretion disk,  irradiated 
by a primary continuum  of  the central engine.
Fabian et
al. (1989) suggested  that the iron line due to  disk reflection 
  is in general broadened and skewed by the Doppler effects and gravitational redshift if the
line is emitted in the vicinity of the black hole. 
In fact, the iron line profile observed with ASCA
is well explained with such a   ``disk-line'' model 
 (Tanaka et al.\ 1995).

Meanwhile, Fabian et al. (2002) and Matsumoto et al. (2003) reported
significant {\em invariability}\/ in the  iron line energy band of MCG
6-30-15 in XMM-Newton and ASCA observations, respectively. 
Inoue and Matsumoto (2003) pointed  out that 
the absorbed spectrum due to photoionized warm absorbers
mimics the shape of the strongly red-shifted iron line,
and that variation of the warm absorbers may explain such 
invariability in the iron line region. 
On the other hand, Miniutti and Fabian (2004) proposed that the observed 
invariability of the iron line may be  explained if
the disk reflection and line photon production 
takes place very close to the black hole, such that the apparent invariability
of the iron line is due to general relativistic light-bending effects.
In fact, Miniutti et al. (2007) claim that the strong reflection,
the broad iron line and  invariability of the iron line
observed with Suzaku may be explained with the light-bending model, 
and suggest that 
the innermost disk radius extends down to about as low as 
two gravitational radii. 
Meanwhile, 
Nied\'{z}wiecki \& \.{Z}ycki (2008) independently re-examined the light-bending model and
concluded that it is indeed possible  
that a reflected component is very weakly variable compared to
the primary emission,  if a source is moving {\em radially}\/ very close to the rotating black hole,
while whether such a particular movement is plausible or not is another question.

On the other hand, a similarly weakly variable, {\em apparently broad}\/ Fe line in
NGC 4151 is most likely to  be an artifact of incorrect modelling of the absorbers,
and the spectrum is in fact dominated by a narrow emission line
 (Ogle et al.\ 2000; de Rosa et al.\ 2007).  
As a matter of  fact, Takahashi, Inoue, and Dotani (2002) 
confirmed, from a very long RXTE monitoring data,  an 
unambiguous evidence that the reverberation of  the iron line
emission takes place   not in the vicinity of the central black hole 
but very far from the black hole ($\sim10^{17}$ cm).
Therefore, at least in NGC 4151, most
iron line photons are produced very far from the black hole and
thus invariable on a timescale of $\sim10^5$ sec.  

In MCG-6-30-15 too,  Young et al. (2005) reported a weak and narrow
emission  line at 6.4 keV, which indicates that some reflection
does take place in a distant material, e.g., the narrow-line region or
the pc-scale torus.
However, equivalent width of the narrow line 
in MCG-6-30-15 ($\sim18$ eV) is much smaller than in NGC 4151 ($\lesssim$ 2 keV; 
Takahashi, Inoue and Dotani 2002).
Therefore, 
invariability of the apparently broadened iron line feature in MCG-6-30-15
has yet to be explained.

The question in MCG-6-30-15 is thus whether the Fe line is broad
and its lack of variability is a result of the relativistic effect, or
it is rather narrow, comes from a distant and extended medium, which would naturally
smear the  variability.  The crucial element to address the problem
is to model the underlying continuum spectrum and study its variation.
The broad band continuum in MCG-6-30-15 is complex, with strong modification 
from warm absorbers (Reynolds et al.\ 1995, Otani et al.\ 1996,
Lee et al.\ 2001, Sako et al.\ 2003, Turner et al.\ 2003, Turner et al.\
2004, Young et al.\ 2005, Miller et al.\ 2008).  
Moreover, the continuum spectral shape is variable, and
the spectral variation is suggested to be  driven mostly by variations of 
the complex warm absorbers (Miller et al.\ 2008).  

In this paper we attempt to comprehensively characterize the spectral 
variability of the source in a model independent manner as much as possible.
We characterize  spectral variations at different timescales, as well as 
at different source flux levels. We use Suzaku data taken in Jan 2006
(total exposure time is 339 ksec) for this purpose.
Suzaku's broad energy coverage (0.6--40 keV),  excellent spectral resolution
and a large effective area, make it an  ideal instrument to study
spectral variation of moderately bright AGNs such as MCG-6-30-15.
Our goal is to find a reasonable spectral model of MCG-6-30-15
 which is able to explain the observed  spectral variation at  various 
timescales with minimum  numbers of parameters, and to study effects of
warm absorbers. 
Such a reasonable spectral model is very needed
to disentangle the tough disk-line problem, 
since the broad iron line feature is affected by the choice of 
continuum spectral models.

\section{Observation and Data Reduction}

The Suzaku satellite (Mitsuda et al. 2007) has observed MCG-6-30-15
($\it z$ = 0.00775; Fisher et al. 1995) four
times. The first observation was performed between
August 17--19 in 2005 for about 45 ksec.
In 2006 January, the source was observed three times, between 9--14 (143 ksec exposure), 
23--26 (99 ksec), and 27--30 (97 ksec).
In this paper, we use the data taken in 2006 January. 

For  data reduction we used the HEADAS 6.5 software package,
provided by NASA/GSFC. 
The XIS data were screened with XSELECT using the  standard criterion (Koyama et
al. 2007). The XIS spectra and light-curves were extracted from circular regions of 3.82 
arcmin radius centred on the source, while background products were
extracted from the outer annulus region with a total area being
equal to that of the source region. 

\begin{figure}[htbp]
\begin{center}
\FigureFile(80mm,80mm){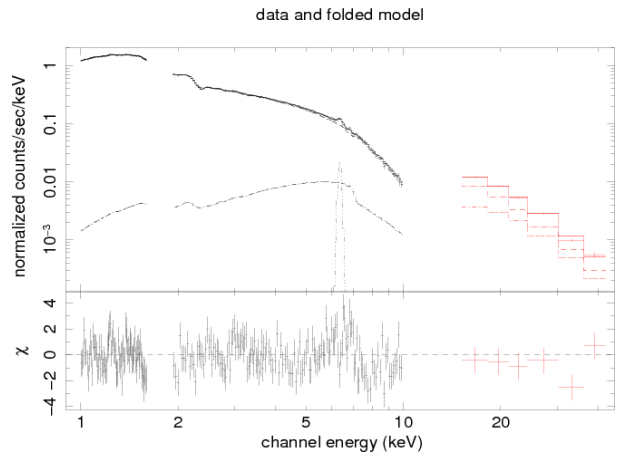}
\end{center}
\caption{Spectral fit result for the time-averaged XIS and PIN spectra
 (1--40 keV) with a {\it narrow}\/ iron emission line
(intrinsic line width is fixed to 1 $\sigma = 10$ eV).  }
 \label{fig1}
\end{figure}

\begin{figure}[htbp]
\begin{center}
\FigureFile(80mm,80mm){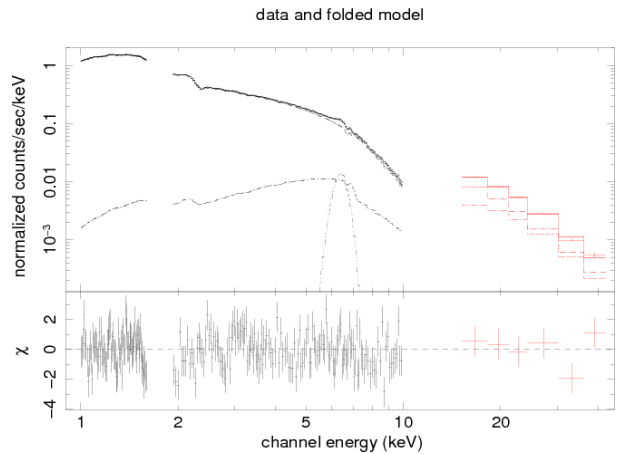}
\end{center}
\caption{Spectral fit result for the time-averaged XIS and PIN spectra
 (1--40 keV) with a {\it broad} iron emission line (intrinsic line width
is allowed to be free and 1 $\sigma$=290 eV.}
\label{fig1-2}
\end{figure}

We generated XIS  response matrices using the {\tt xisrmfgen} software, which  takes into account
the time-variation of the energy response. 
As for generating ancillary response files (ARFs), we used
{\tt xissimarfgen}  (Ishisaki et al. 2007). This tool calculates ARFs
through  ray-tracing, and we selected the number of input photons as 400000, 
with the ``estepfile'' parameter ``full''.

After extracting the products for the back-illuminated CCD (XIS1) 
and for the three front-illuminated CCD XIS detectors (XIS0, XIS2,
XIS3) separately, we found the XIS0, XIS2, and XIS3 products are almost
identical to each other. So we combined the
XIS0, XIS2, and XIS3 products in our analysis. We used the {\tt addascaspec} 
to combine the XIS spectra and responses.

As for HXD/PIN (Takahashi et al. 2007), instrumental background model  is
provided by the HXD instrument team  in the form of  simulated PIN event files. 
These background event files do not contain the cosmic
X-ray background (CXB) component, 
which we  estimated separately using 
the PIN response for flat emission
distribution ({\tt ae$\_$hxd$\_$pinflate1$\_$20080129.rsp}) assuming the CXB spectrum
measured by HEAO-1   (Boldt 1987). 

\section{Data Analysis and Results}
\subsection{Spectral Fitting for the  Average Spectrum}
First, in this section, we study  the  time-average spectrum of MCG-6-30-15
to find a physically plausible spectral model. Then, 
in the following sections,  we will use the same spectral model 
to see which parameters are variable to explain the observed spectral variation (sections \ref{sec:slicespectra} and \ref{sec:brightfaint}).
The spectral  model we adopt
is the following;  (1) power-law with an exponential cut-off,
(2) disk reflection component from neutral matter (``pexrav''; Magdziarz $\&$
Zdziarski 1995), 
(3) iron emission line,
(4) two warm absorbers with different ionization states, and
(5) neutral photoelectric absorption  (``phabs''; Balucinska-Church $\&$ McCammon 1992). 
(6) a narrow gaussian absorption line to account for the instrumental feature
around the Au M-edge.

We use  XSTAR Version 2.1kn8 (Kallman et al. 2004; see also the note in the end of the paper) 
to model the warm absorbers, assuming 
 the solar abundance and the photon index of the ionizing spectrum  2.0.
The temperature, pressure and density of the warm
absorbers are assumed to be  ${10}^5$ K,  0.03 dyne/${\rm cm}^2$ and ${10}^{12} {\rm cm}^{-3}$, respectively.  We made a grid model 
  by running 
XSTAR for different values of $\xi$ and $N_H$; 
the log $\xi$ values are from 
0.1 to  5 (erg cm s$^{-1}$)  and
$N_H$ values are from 
${10}^{20}$ to  ${10}^{24}$ (cm$^{-2}$). The number of
steps for log $\xi$ and $N_H$ are both 20, thus our  grid model has
$20 \times 20$ grid-points. 

All the fits were made with XSPEC v11 (Arnaud 1996). 
In  Fig.$\ref{fig1}$, we show the fitting result
for total average spectrum with a {\it narrow}\/ emission line
(line width is fixed at $1\sigma=10$ eV), where  reduced chisq is 1.51 (${\chi}^2$/d.o.f =
339.3/224). 
The broadband 1--40 keV spectrum was fitted reasonably well, while
there remain some residuals in the energy range of 5--7
keV (within $\pm$ 5 \%)  for putatively broad iron emission line.
Note that the XSTAR warm absorber model includes the absorption lines,
and in fact the observed Fe{\footnotesize XXV} and Fe{\footnotesize XXVI} absorption line features at
6.67 keV and 6.97 keV are successfully modeled.
However, we found this  spectral model fails to fit the data below 1 keV; this issue is discussed separately in section \ref{Below1keV}.

\begin{figure}[htbp]
\begin{center}
\FigureFile(80mm,50mm){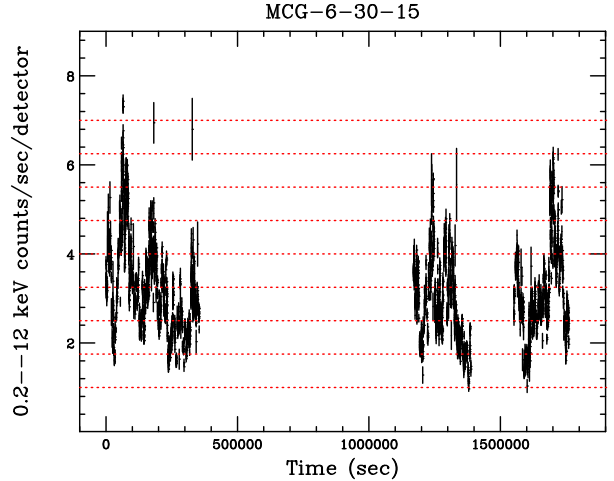}
\caption{0.2--12 keV XIS light curve of the 2006 observation. 
The  count rate  intervals with which intensity-sliced spectra are made are also indicated.
} 
\label{lc}
\end{center}
\end{figure}

If we allow the intrinsic line width to be free, 
the fit significantly improved, where  reduced chisq is 1.20
(${\chi}^2$/d.o.f = 267.4/222).  Central energy of the line is
6.42$\pm$0.06 keV, and the width is
1 $\sigma = 290 \pm 60$ eV (Figure \ref{fig1-2}).  The equivalent width is
100$\pm20 $ eV. 
In  Fig.$\ref{fig1-2}$, we show the fitting result for the total average
spectrum with a broad line.  Note that our model  does not require an
extremely broadened iron emission line, which may be expected from very vicinity of
a fast rotating black hole. The best-fit parameters are shown in Table$\ref{table0}$.

\begin{table}
\caption{Results of the average spectral fitting in 1--40 keV\footnotemark[$*$]}
\begin{tabular}{lccc}
\hline
                                    &broad line & narrow line\\
\hline
$N_H$ (${10}^{21}$ ${\rm cm}^{-2}$)  &  2.2$\pm$0.2 &  1.7$\pm$0.2 \\
\hline
$N_H$  (${10}^{22}$ ${\rm cm}^{-2}$)  & 6$^{+3}_{-2}$ & 10$^{+6}_{-4}$ \\
log $\xi$  & 3.3$\pm$0.1  & 3.6$\pm$0.2  \\
\hline
$N_H$  (${10}^{21}$ ${\rm cm}^{-2}$)  & 2.4$\pm$0.8 &
		 1.5$\pm$0.4 \\
log $\xi$  & 1.7$^{+0.1}_{-0.3}$ & 1.0$^{+0.2}_{-0.4}$ \\
\hline
Line E (keV) & 6.42$\pm$0.06 &  6.35 (fix) \\
sigma (keV) & 0.29$\pm$0.06 & 0.01 (fix)  \\
norm (${10}^{-5}$) & 4.4$\pm$0.7 & 1.7$\pm$0.3  \\
EW (eV) &  100$\pm$20  & 38$\pm$6 \\
\hline
cutoffpl K (10$^{-2}$) & 1.47$\pm$0.05 & 1.38$\pm$0.03 \\
photon index & 1.95$\pm$0.02 & 1.90$\pm$0.02\\
\hline
$E_{cut}$ (keV) & 160 (fix) & 160 (fix) \\
cosIncl & 0.866 (fix) & 0.866 (fix)  \\
$\Omega$/2$\pi$  & 1.0$\pm$0.2 & 0.8$\pm$0.2 \\
\hline
Line E (keV) & 2.35$\pm$0.02 & 2.35$\pm$0.02 \\
sigma (keV) & 0.01(fix) & 0.01 (fix) \\
norm (${10}^{-5}$) & -2.2$\pm$0.5 & -2.2$\pm$0.5 \\
\hline
reduced chisq (d.o.f) & 1.20 (222) & 1.51 (224) \\
\hline
\multicolumn{3}{@{}l@{}} {\hbox to 0pt {\parbox{85mm}{\footnotesize
\footnotemark[$*$] Errors are quoted at statistical 90$\%$ level. 
 Constant factor to
adjust normalizations between XIS and PIN is 1:1.086 (Ishida, Suzuki, and Someya
 2007). Units of the cutoff power-law and pexrav normalization are
photons/s/cm$^2$/keV at 1 keV.
$\Omega$ is solid angle of the reflector seen from the
central source, and $\Omega$/2$\pi$ is defined as the ratio of the 
normalization of the reflection component to that of  the cutoff power-law component.
 Unit of the line normalization is photons/s/cm$^2$.
Redshift of warm absorber and objects are 0.001 and 0.00775,
 respectively (Young et al. 2005, Fisher et al. 1995).
 Energy cutoff ($E_{cut}$) is fixed to 160 keV (Guainazzi et
 al. 1999). The solar abundances are defined in Greeves, Noels and
 Sauval (1996) for the warm absorbers and  in Anders \& Ebihara (1982) for 
the reflection component (pexrav). 
 }\hss}}
\label{table0}
\end{tabular}
\end{table}

\subsection{Spectral Variation}
\label{sec:slicespectra}

Next, we study spectral variation of the source in a model-independent manner.
First, we make ``intensity-sliced energy  spectra'' as follows:

\noindent
1) Create a light curve (the average of XIS0, XIS2, and XIS3), with a bin-width of 128 sec in the 0.2 -- 12 keV
band.  We found that the count rate varies  from $\sim 1$ to $\sim 7$
cnts/sec. Figure$\ref{lc}$ shows the 0.2--12 keV XIS light curve 
 used in the present data analysis.

\noindent
2) Choose time periods when the source intensity is in the ranges of  1--1.75, 1.75--2.50,
2.50--3.25, 3.25--4.00, 4.00--4.75, 4.75--5.50, 5.50--6.25, 6.25--7.00 cnts/sec.
These intensity ranges are chosen so that the exposure times for individual intensity bins
are approximately equal.

\noindent
3) From the eight intervals corresponding to the different source flux levels,
we create eight intensity-sliced energy spectra (the sum of XIS0, XIS2,
and XIS3).

Second, to study spectral variations in various timescales,
we extract ``bright spectra'' and ``faint spectra''  as follows: 

\noindent 1) Divide
the entire observation period into a series of the time-interval of 
which length is $T$.  

\noindent  2) Create light curves for individual intervals
with a time-bin-width of 128 sec,
and calculate the average count rate for each interval $T$. 

\noindent 3) For each interval, create 
the ``bright spectrum'' from the period when the XIS count rates 
(0.2 -- 12 keV)  are higher
than the average, and the ``faint spectrum'' when the count rates are
lower  than the average. 

\noindent  4) Average  the bright spectra and faint spectra for
 all the intervals. 

\noindent 5) Thus, for a given $T$,  we have one 
 bright spectrum  and one faint spectrum.

\noindent 6) Repeat the procedure
for different time-intervals  of $T$.
As for $T$, we chose 5,000, 9,000, 15,000, 40,000, 75,000 and 200,000 sec.

Thus, we have six ``bright  spectra'' and six ``faint spectra'' corresponding to these timescales. 
Note that there are data gaps within a single time interval,
since Suzaku has a low-earth orbit and the observation is intermittent every
$\sim3,000$ seconds or so.
Typical  exposure is $\sim50 \%$ of the bin-size.

\begin{figure}[htbp]
\begin{center}
\FigureFile(100mm,100mm){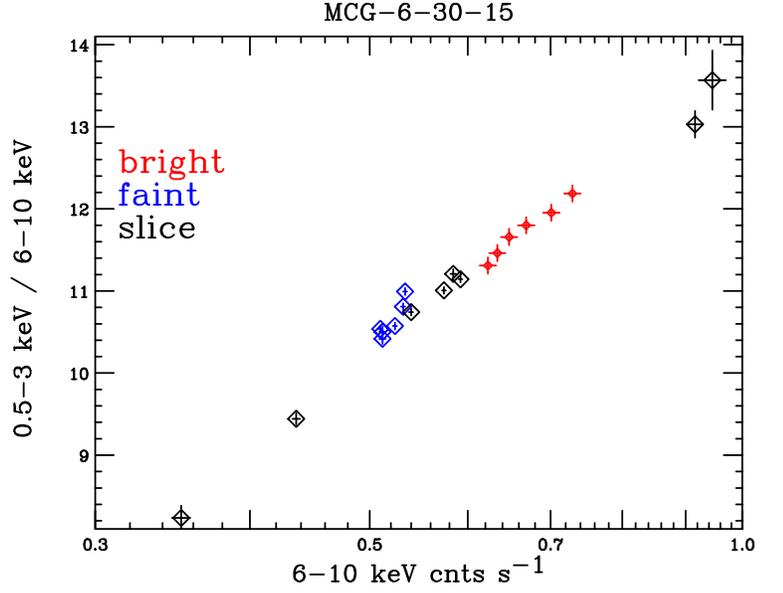}
\caption{Correlation between the flux in 6.0--10 keV and the hardness ratio of 0.5--3.0 keV/6.0--10 keV
for the ``intensity-sliced spectra'' (black), ``bright spectra'' (red) and ``faint spectra'' (blue).}
\label{fig:intensityhardness}
\end{center}
\end{figure}


From the eight intensity-sliced spectra, six bright spectra and six faint spectra thus created, 
we studied  spectral variations in a model independent manner
by calculating spectral hardness radios between different energy bands.
Consequently, we discovered a significant correlation between the
intensity in the 6.0 -- 10 keV and the spectral ratio of 0.5 -- 3.0 keV
/ 6.0 -- 10 keV.
  This correlation is shown in Figure \ref{fig:intensityhardness}.

\begin{figure}[htbp]
\begin{center}
\FigureFile(80mm,80mm){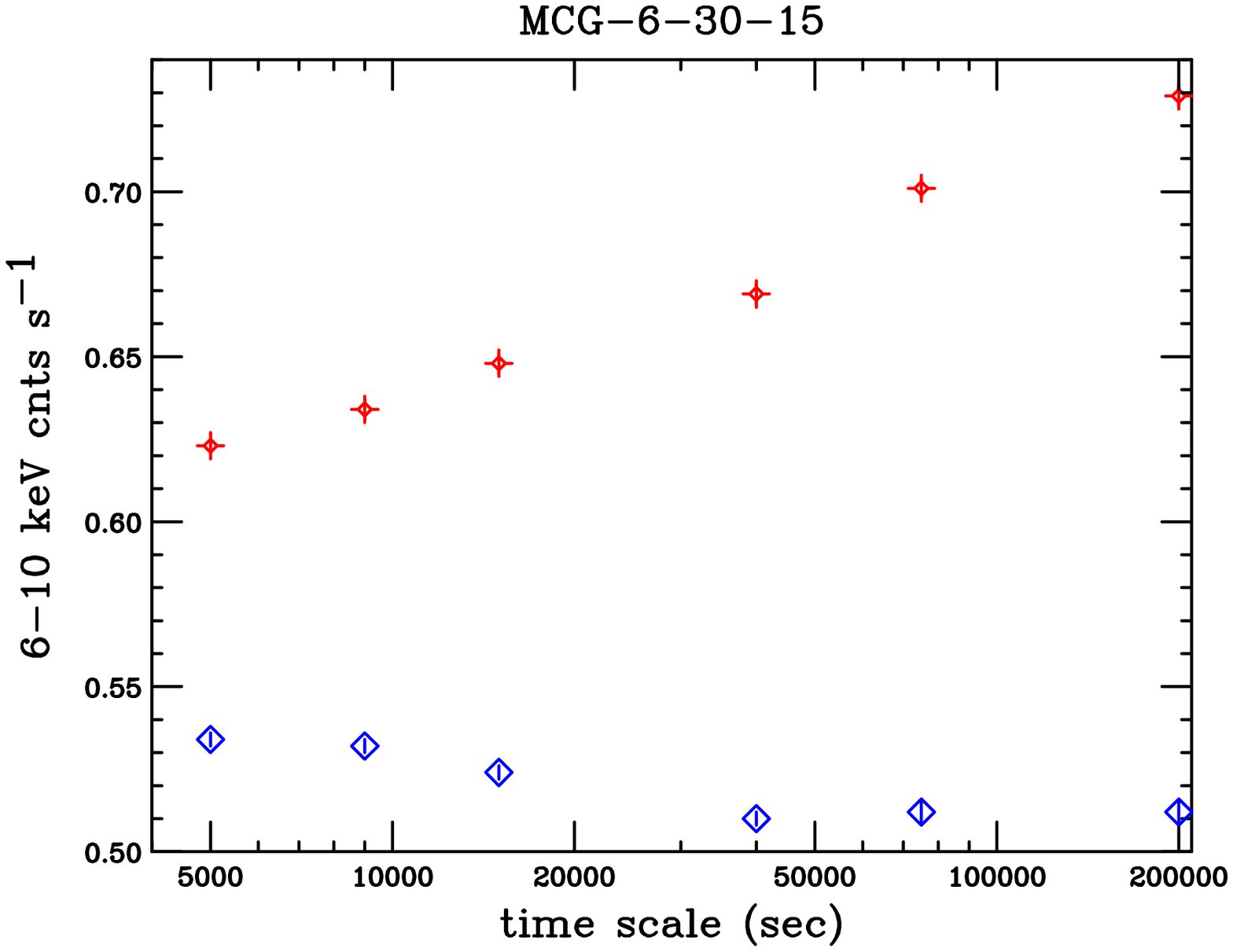}
\FigureFile(80mm,80mm){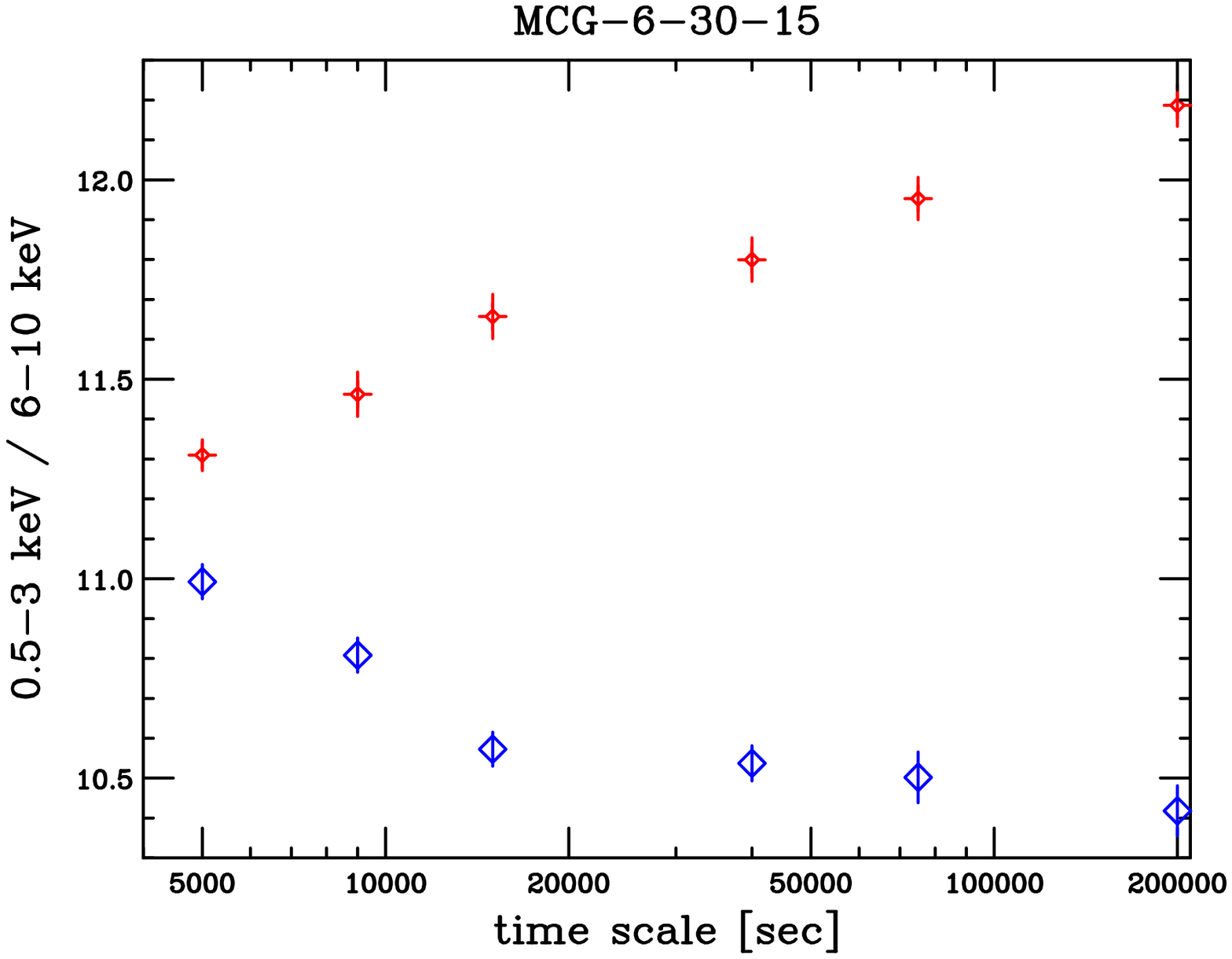}
\end{center}
\caption{The 6.0 -- 10 keV intensity (top) and the spectral hardness ratio (bottom)
of the bright (red) and faint (blue) 
spectra as functions of the time-intervals, $T$.
 }
 \label{data345}
\end{figure}

\begin{figure}[htbp]
\begin{center}
\FigureFile(80mm,80mm){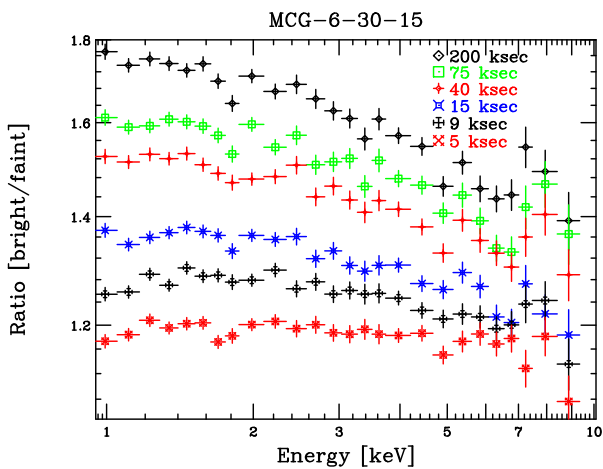}
\end{center}
\caption{Spectral ratios of the bright spectra  to the faint spectra  for six
different  timescales.}
 \label{data6}
\end{figure}

Figure \ref{data345} gives the 6.0 -- 10 keV intensity and the 0.5 --3.0
/ 6.0 -- 10 keV spectral-ratio
as functions of the time-interval $T$.  
From these figures, it is obvious that the intensity and spectral variations are
more prominent with increasing timescales.
In fact, Figure \ref{data6} shows ratios of the bright spectra to the  faint spectra for
different time-intervals.  It is obvious that the ``bright''  spectra are steeper
than the ``dim'' spectra below $\sim$10 keV for all the time scales, 
and that the spectral difference 
becomes  more significant  with increasing  timescales.
We emphasize that the spectral variation we found here 
is {\em model independent}.

\begin{figure}[htbp]
\begin{center}
\FigureFile(80mm,80mm){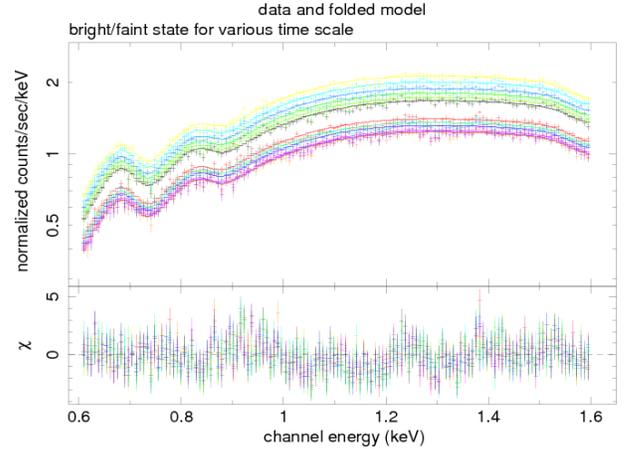}
\end{center}
\caption{Simultaneous fit in the 0.6--1.6 keV band for the twelve spectra 
(bright and faint spectra for  the 6 timescales) with a power-law plus two absorption edge model. }\label{below1keV}
\end{figure}

\subsection{Spectral Fitting for the  Bright and Faint State  Spectra}
\label{sec:brightfaint}
Next, we  quantify  the observed spectral variation on different 
timescales using the same model as for the average spectrum.
We try to fit the six ``bright''  spectra and the 
six ``faint'' spectra simultaneously (both XIS and PIN
in 1 -- 40 keV) with minimum numbers of variable parameters.
As a result, we found that the twelve spectra
are fitted reasonably well with keeping the reflection component constant, and
only varying  normalization of the direct component and another  parameter to
describe its apparent spectral slope, such that the direct component spectrum steepens as the flux
increases (Figures \ref{fig:intensityhardness} and  \ref{data6}).
Namely, we found {\em only two parameters are required}\/ to describe
continuum spectral variations  of the twelve data-sets, whereas all the other spectral 
parameters are constant (iron line flux is slightly variable, which hardly affects
the overall spectral shape).
We found that either power-law photon index, ionization parameter or column density of the lower-ionized warm absorber 
almost equally reproduces the observed spectral change of the direct component.
The best-fit parameters are shown in Table $\ref{table2}$ 
for the twelve spectra,
when the power-law index, the ionization parameter and the column density is varied,
respectively, besides the power-law normalization.

\subsection{Spectral Variation below 1 keV}\label{Below1keV}

We found our two warm-absorber model fails to explain the energy spectra
below 1 keV. Lee et al. (2001) reported strong  oxygen absorption lines
 using Chandra grating observation, 
which suggests there is another warm absorber component responsible for the
oxygen features below 1 keV.
In any case, our aim is to explain and quantify
the observed spectral variation with a simple model as much as 
possible, thus we fit the 0.6--1.6 keV spectra using a
power-law component and two absorption edges to approximate the ionized oxygen edges and
absorption lines.
The twelve spectra are fitted simultaneously, as in the previous section.

The best-fit absorption edge energies are 0.706 keV and 0.855 keV,
which are considered to be  primarily  due to absorption edges/lines of
O{\footnotesize VII} and O{\footnotesize VIII}, respectively.
We found that we can fit all the twelve spectra with common photon index and  the 0.706 keV edge optical depth,
while only the power-law normalization and the optical depth of the 0.855 keV 
edge are variable (Fig. \ref{below1keV}).  Reduced chisq was 1.30 (${\chi}^2$/d.o.f = 2077.7/1604).

The best-fit  parameters are shown in Table $\ref{table4}$. We found that
optical depth of the O VIII edge becomes deeper  and shallower for the
faint state and the  bright state, respectively,
as the  variation timescale increases  (bottom panel in Fig. $\ref{below1keVvariation}$).
This result suggests that the ionization degree of the warm absorber is varying, since 
the oxygen is considered to be more highly ionized and the
O{\footnotesize VIII} edge depth becomes shallower,  as the
source brightens and the warm absorber  gets more highly ionized.  This is consistent
with what was observed with ASCA (Otani et al.\ 1996).
Invariable  O{\footnotesize VII} depth regardless of the flux variation, which was
 also  observed with  ASCA (Otani et al.\ 1996), 
suggests that the O{\footnotesize VII} edge originate from a rather extended 
area in the line of sight, so 
that the O{\footnotesize VII} edge depth does not respond to the flux variation 
instantaneously.
The point here is that the spectral variation below 1.6 keV 
can be reproduced by sole  variation of the photoionization as a natural consequence of the
intrinsic flux changes, while the
 photon index is invariable.


\setlength{\tabcolsep}{2.4pt}
\begin{table*}
\begin{center}
\rotatebox{90}{
\begin{minipage}[t]{\textwidth}
\scriptsize
\caption{Results of spectral fitting in 1--40 keV for the ``bright'' and
 ``faint'' spectra.  \footnotemark[$*$]} 
\begin{tabular}{ccccccccccccc}
\hline
timescale & \multicolumn{2}{c} {5 ksec} & \multicolumn{2}{c}{9 ksec} & \multicolumn{2}{c}{15 ksec} & \multicolumn{2}{c} {40 ksec} & \multicolumn{2}{c} {75ksec} & \multicolumn{2}{c} {200 ksec} \\
\hline
state  & bright & faint & bright & faint & bright & faint & bright & faint & bright & faint & bright & faint  \\
\hline\hline
\multicolumn{13}{l}{\it Model 1: power-law normalization and its index  are variable} \\
\hline
$N_H$ (${10}^{21}$ ${cm}^{-2}$)  &
	 \multicolumn{12}{c}{1.7$\pm$0.1} \\
\hline
$N_H$  (${10}^{22}$ ${cm}^{-2}$)  &
	 \multicolumn{12}{c}{10$\pm$2} \\
log $\xi$  &  \multicolumn{12}{c}{3.56$\pm$0.05} \\
\hline
$N_H$  (${10}^{21}$ ${cm}^{-2}$)  & \multicolumn{12}{c}{1.4$\pm$0.2} \\
log $\xi$   & \multicolumn{12}{c}{1.03$^{+0.03}_{-0.11}$} \\
\hline
Line E (keV) & \multicolumn{12}{c}{6.35 (fix)} \\
sigma (keV) & \multicolumn{12}{c}{0.01 (fix)} \\
norm (${10}^{-5}$) & \multicolumn{12}{c}{ 1.7$\pm$0.1} \\
EW (eV) &  35$\pm$2  &  41$\pm$3  &   35$\pm$2  &  42$\pm$3  &
					 34$\pm$2  & 42$\pm$3 &   33$\pm$2   &  44$\pm$3   &   31$\pm$2  & 44$\pm$3    & 30$\pm$2 &  44$\pm$3  \\
\hline
cutoffpl K (10$^{-2}$) &   1.51$^{+0.01}_{-0.02}$  & 1.26$^{+0.01}_{-0.02}$  & 1.57$^{+0.01}_{-0.02}$  &
				 1.22$^{+0.01}_{-0.02}$ &    1.64$^{+0.01}_{-0.02}$  & 1.18$^{+0.01}_{-0.02}$ &
 1.72$^{+0.01}_{-0.02}$  &  1.11$^{+0.01}_{-0.02}$ & 1.83$^{+0.01}_{-0.03}$ &  1.13$^{+0.01}_{-0.02}$ &  1.95$\pm$0.02 &
 1.10$^{+0.01}_{-0.02}$ \\
photon index &   1.91$\pm$0.01 &  1.90$\pm$0.01 &   1.91$\pm$0.01  &  1.89$\pm$0.01 &   1.92$\pm$0.01    &    1.88$\pm$0.01  &  1.93$\pm$0.01   &  1.88$\pm$0.01   &  1.94$\pm$0.01   & 1.88$\pm$0.01    &  1.95$\pm$0.01   &  1.87$\pm$0.01    \\
\hline
$E_{cut}$ (keV) & \multicolumn{12}{c}{160 (fix)} \\
cosIncl & \multicolumn{12}{c}{0.866 (fix)} \\
pexrav K (10$^{-2}$)  & \multicolumn{12}{c}{1.07$\pm$0.08} \\
\hline
Line E (keV) & \multicolumn{12}{c}{2.35$\pm$0.01} \\
sigma (keV) & \multicolumn{12}{c}{0.01 (fix)} \\
norm (${10}^{-5}$) & \multicolumn{12}{c}{-2.12$\pm$0.22} \\
\hline
reduced chisq (d.o.f) & \multicolumn{12}{c}{1.30 (2786)} \\
\hline\hline
\multicolumn{13}{l}{\it Model 2: power-law
 normalization and  ionization degree of the lower-ionized warm absorber are variable} \\
\hline
$N_H$ (${10}^{21}$ ${cm}^{-2}$)  &  
		 \multicolumn{12}{c}{2.14$\pm$0.05} \\
\hline
$N_H$  (${10}^{22}$ ${cm}^{-2}$)  &
		 \multicolumn{12}{c}{3.2$^{+0.8}_{-0.6}$} \\
log $\xi$  & \multicolumn{12}{c}{3.29$\pm$0.04} \\
\hline
$N_H$  (${10}^{21}$ ${cm}^{-2}$)  & 
		 \multicolumn{12}{c}{2.5$\pm$0.3} \\
log $\xi$  &  1.57$\pm$0.05  &
			 1.56$^{+0.05}_{-0.06}$ &  1.60$\pm$0.05   &
					 1.53$^{+0.05}_{-0.06}$  &  1.67$\pm$0.05 &
							 1.46$^{+0.05}_{-0.09}$ & 1.74$^{+0.05}_{-0.06}$  & 1.41$\pm$0.06  & 1.80$^{+0.05}_{-0.06}$ &   1.43$\pm$0.06  &  1.86$^{+0.07}_{-0.06}$     &   1.39$^{+0.05}_{-0.06}$        \\
\hline
Line E (keV) & \multicolumn{12}{c}{6.35 (fix)} \\
sigma (keV) &  \multicolumn{12}{c}{0.01 (fix)} \\
norm (${10}^{-5}$) & \multicolumn{12}{c}{1.7$\pm$0.1} \\
EW (eV) & 36$\pm$2  & 42$\pm$3 & 35$\pm$2  & 43$\pm$3  & 34$\pm$2  &  43$\pm$3  &  33$\pm$2  &  45$\pm$3 & 31$\pm$2  & 45$\pm$3   &  30$\pm$2     & 46$\pm$3 \\
\hline
cutoffpl K (10$^{-2}$) & 1.64$\pm$0.02 &
				 1.37$\pm$0.02 & 1.68$^{+0.02}_{-0.01}$ &
						 1.33$\pm$0.02 &  1.73$\pm$0.02 & 1.31$\pm$0.02  &  1.79$\pm$0.02  & 1.25$\pm$0.01 & 1.90$\pm$0.02 &  1.26$\pm$0.01  & 2.00$\pm$0.02  &  1.24$\pm$0.01 \\
photon index & 
		 \multicolumn{12}{c}{1.96$\pm$0.01} \\
\hline
$E_{cut}$ (keV) & \multicolumn{12}{c}{160 (fix)} \\
cosIncl & \multicolumn{12}{c}{0.866 (fix)} \\
pexrav K (10$^{-2}$)  & 
		 \multicolumn{12}{c}{1.7$\pm$0.1} \\
\hline
Line E (keV) &  \multicolumn{12}{c}{2.35$\pm$0.01} \\
sigma (keV) & \multicolumn{12}{c}{0.01 (fix)} \\
norm (${10}^{-5}$) & \multicolumn{12}{c}{-2.3$\pm$0.2} \\
\hline
reduced chisq (d.o.f) & \multicolumn{12}{c}{1.41 (2787)} \\
\hline\hline
\multicolumn{13}{l}{\it Model 3: power-law
 normalization and column density of the lower-ionized warm absorber  are variable} \\
\hline
$N_H$ (${10}^{21}$ ${cm}^{-2}$)  &
	 \multicolumn{12}{c}{2.2$\pm$0.1} \\
\hline
$N_H$  (${10}^{22}$ ${cm}^{-2}$)  &
	 \multicolumn{12}{c}{3.0$\pm$0.7} \\
log $\xi$  &  \multicolumn{12}{c}{3.28$\pm$0.04} \\
\hline
$N_H$  (${10}^{21}$ ${cm}^{-2}$)  &    2.8$\pm$0.4  &
		 3.0$\pm$0.4   & 2.5$\pm$0.4	 &
				 3.2$\pm$0.4   &  2.1$\pm$0.4   &
						 3.7$\pm$0.4 &  1.8$\pm$0.3  & 4.0$\pm$0.5  & 1.4$\pm$0.4  &  3.9$^{+0.5}_{-0.4}$   &  1.1$\pm$0.3  &  4.2$^{+0.5}_{-0.4}$        \\
log $\xi$  &  \multicolumn{12}{c}{1.73$\pm$0.05} \\
\hline
Line E (keV) & \multicolumn{12}{c}{6.35 (fix)} \\
sigma (keV) & \multicolumn{12}{c}{0.01 (fix)} \\
norm (${10}^{-5}$) & \multicolumn{12}{c}{  1.7$\pm$0.1} \\
EW (eV) &  36$\pm$2  & 42$\pm$3   &  35$\pm$2  &  43$\pm$3  &  34$\pm$2  &   43$\pm$3  & 33$\pm$2 & 45$\pm$3 & 32$\pm$2  & 45$\pm$3 & 30$\pm$2 & 46$\pm$3  \\ 
\hline
cutoffpl K (10$^{-2}$) &      1.63$^{+0.03}_{-0.02}$ & 1.37$\pm$0.02 & 1.67$^{+0.03}_{-0.02}$   &  1.33$\pm$0.02 & 1.71$^{+0.03}_{-0.02}$    & 1.31$\pm$0.02  &   1.77$^{+0.03}_{-0.02}$  & 1.25$\pm$0.02   & 1.87$^{+0.03}_{-0.02}$  &  1.26$^{+0.02}_{-0.01}$  & 1.96$^{+0.03}_{-0.02}$    &       1.24$\pm$0.02     \\
photon index &  \multicolumn{12}{c}{1.96$\pm$0.01}    \\
\hline
$E_{cut}$ (keV) & \multicolumn{12}{c}{160 (fix)} \\
cosIncl & \multicolumn{12}{c}{0.866 (fix)} \\
pexrav K (10$^{-2}$)  & \multicolumn{12}{c}{1.56$\pm$0.01} \\
\hline
Line E (keV) & \multicolumn{12}{c}{2.35$\pm$0.01} \\
sigma (keV) & \multicolumn{12}{c}{0.01 (fix)} \\
norm (${10}^{-5}$) & \multicolumn{12}{c}{-2.26$\pm$0.23} \\
\hline
reduced chisq (d.o.f) & \multicolumn{12}{c}{1.35 (2787)} \\
\hline
\multicolumn{13}{@{}l@{}} {\hbox to 0pt {\parbox{185mm}{\footnotesize
\footnotemark[$*$] 
See the footnote  of Table \protect\ref{table0}.
}\hss}}
\label{table2}
\end{tabular}
\end{minipage}}
\end{center}
\end{table*}

\setlength{\tabcolsep}{2.4pt}
\begin{table*}
\begin{center}
\rotatebox{90}{
\begin{minipage}[t]{\textwidth}
\scriptsize
\caption{Table 3. Results of spectral fitting in 0.6--1.6  keV for the ``bright'' and
``faint'' spectra. \footnotemark[$*$] }
\begin{tabular}{ccccccccccccc}
 \hline
timescale & \multicolumn{2}{c} {5 ksec} & \multicolumn{2}{c} {9 ksec} & \multicolumn{2}{c}{15 ksec}  & \multicolumn{2}{c} {40 ksec} & \multicolumn{2}{c} {75 ksec} & \multicolumn{2}{c}{200 ksec} \\
\hline
state  & bright & faint & bright & faint & bright & faint  & bright & faint & bright & faint & bright & faint  \\
\hline
photon index & \multicolumn{12}{c}{1.82$\pm$0.05} \\
K (${10}^{-2}$) &  1.28$\pm$0.01  &  1.06$\pm$0.01 & 1.31$\pm$0.01 &
				 1.03$\pm$0.01 & 1.37$\pm$0.01  &  1.00$\pm$0.01 &
 1.43$\pm$0.01 &  0.95$\pm$0.01 & 1.52$\pm$0.01 & 0.96$\pm$0.01
 &   1.60$\pm$0.01  &   0.94$\pm$0.01 \\
\hline
 edge E (keV) & \multicolumn{12}{c}{0.706$\pm$0.001} \\
MaxTau & \multicolumn{12}{c}{0.67$\pm$0.01} \\
\hline
edge E (keV) & \multicolumn{12}{c}{0.855$\pm$0.002} \\
MaxTau & 0.34$\pm$0.02 & 0.31$\pm$0.02 & 0.33$\pm$0.02 & 0.32$\pm$0.02 & 0.33$\pm$0.02 & 0.32$\pm$0.02  & 0.31$\pm$0.02 & 0.34$\pm$0.02 &  0.32$\pm$0.02 & 0.35$\pm$0.02 & 0.30$\pm$0.02 & 0.34$\pm$0.02 \\
\hline
reduced chisq (d.o.f) & \multicolumn{12}{c}{1.30 (1604)} \\
\hline
\multicolumn{13}{@{}l@{}} {\hbox to 0pt {\parbox{185mm}{\footnotesize
\footnotemark[$*$] Errors are quoted at statistical 90$\%$ level.}\hss}} 
\label{table4}
\end{tabular}
\end{minipage}}
\end{center}
\end{table*}

\section{Discussion}

\begin{figure}
\begin{center}
\FigureFile(80mm,80mm){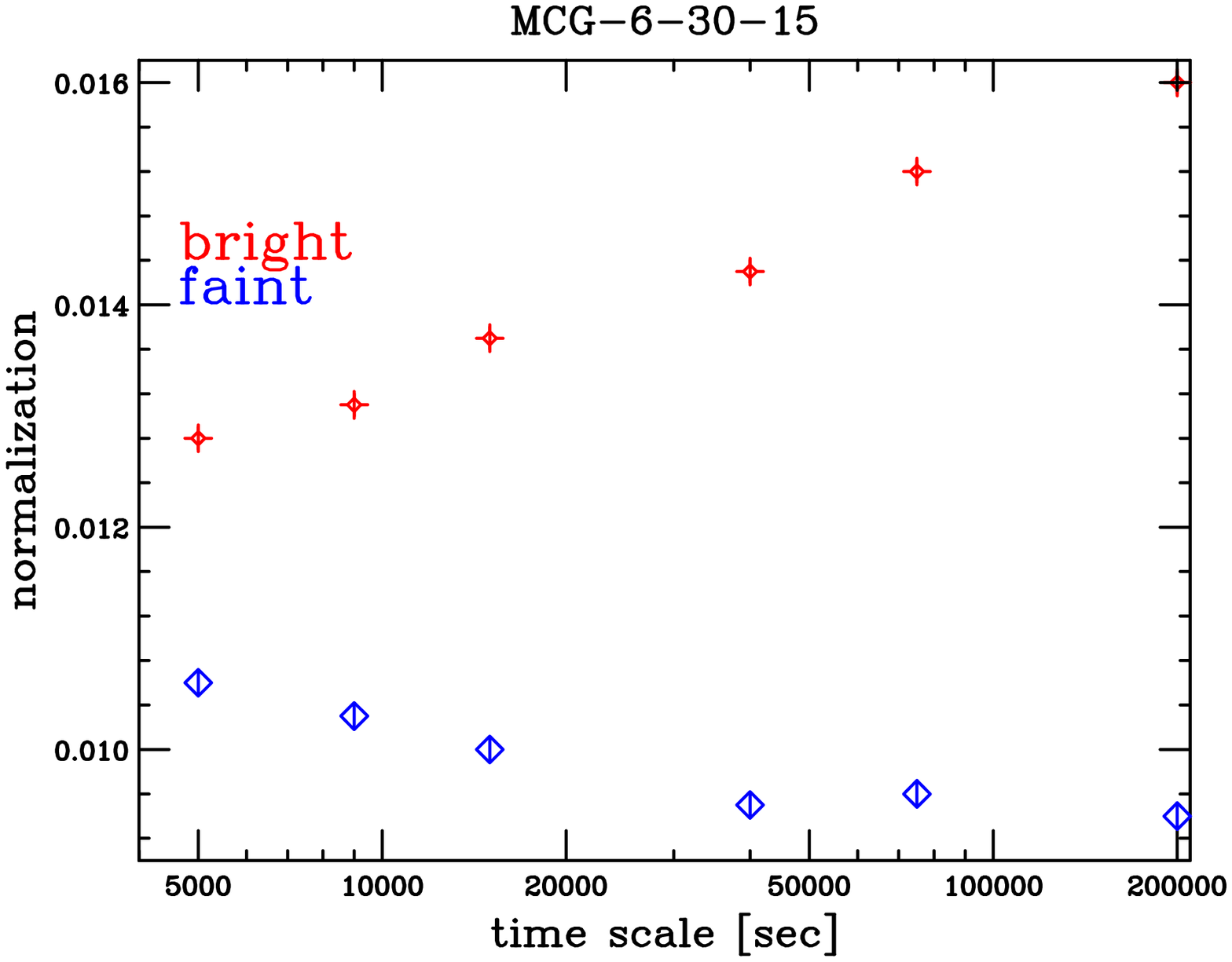}
\FigureFile(80mm,80mm){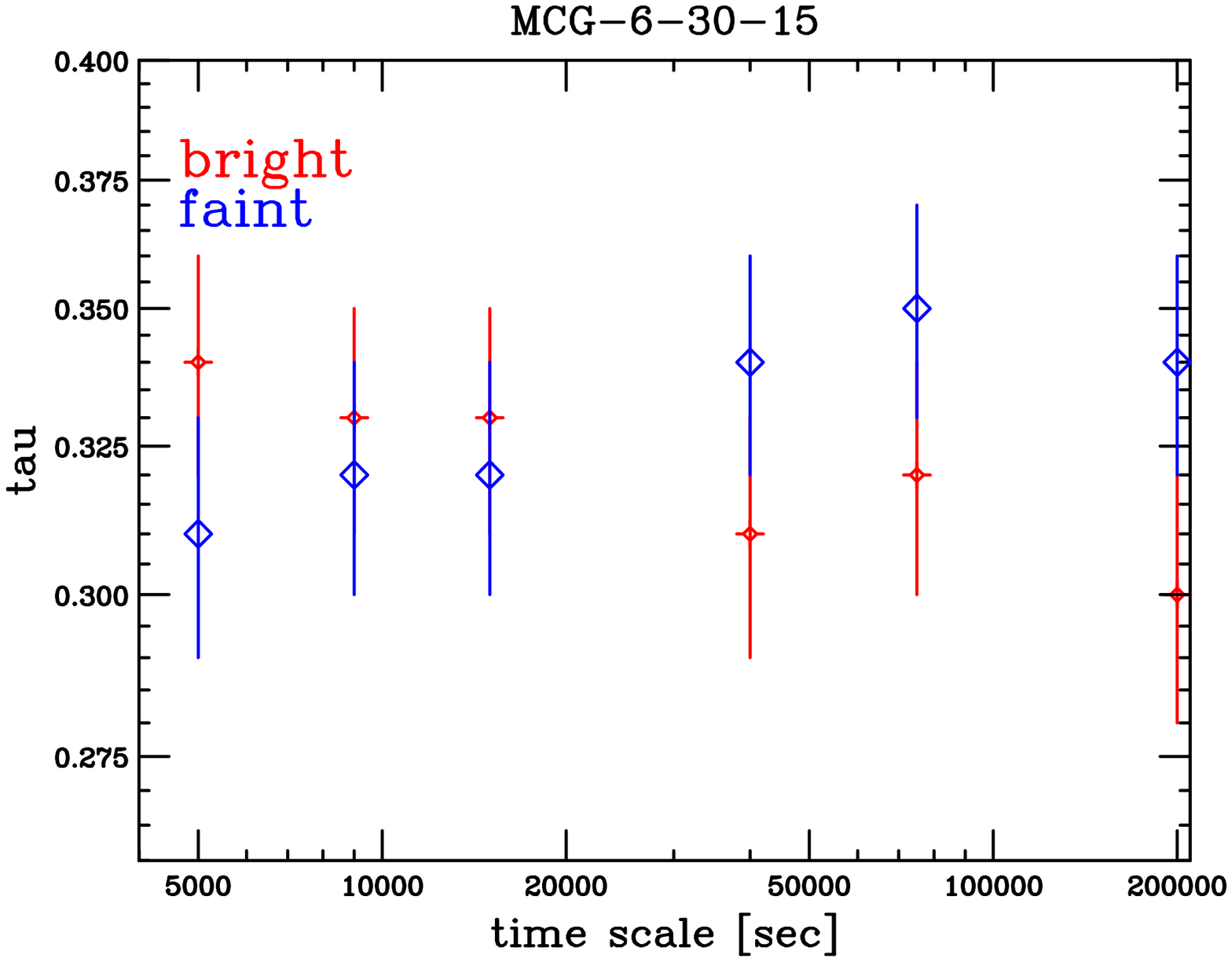}
\end{center}
\caption{Variation of the power-law normalization and depth of the O{\footnotesize VIII} edge 
to describe the spectral variation in  0.6 -- 1.6 keV, as functions of the timescales.}
\label{below1keVvariation}
\end{figure}

\begin{figure}[htbp]
\begin{center}
\FigureFile(80mm,80mm){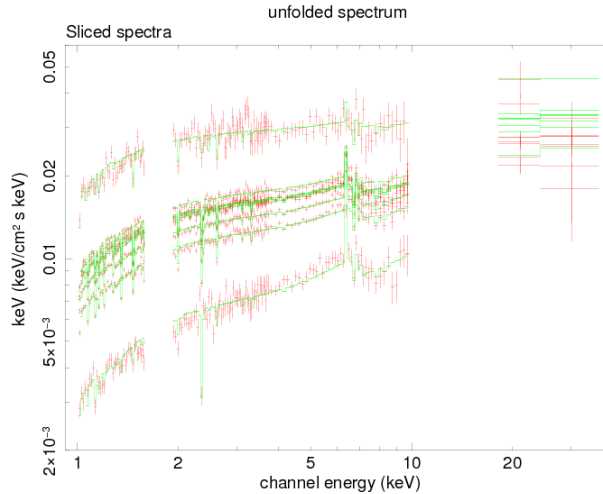}
\end{center}
\caption{Unfolded energy spectra for the eight intensity-sliced spectra
calculated from the model only varying the power-law normalization and
ionization degree of the lower-ionized warm absorber. }
 \label{data12}
\end{figure}

\begin{figure}[htbp]
\begin{center}
\FigureFile(80mm,80mm){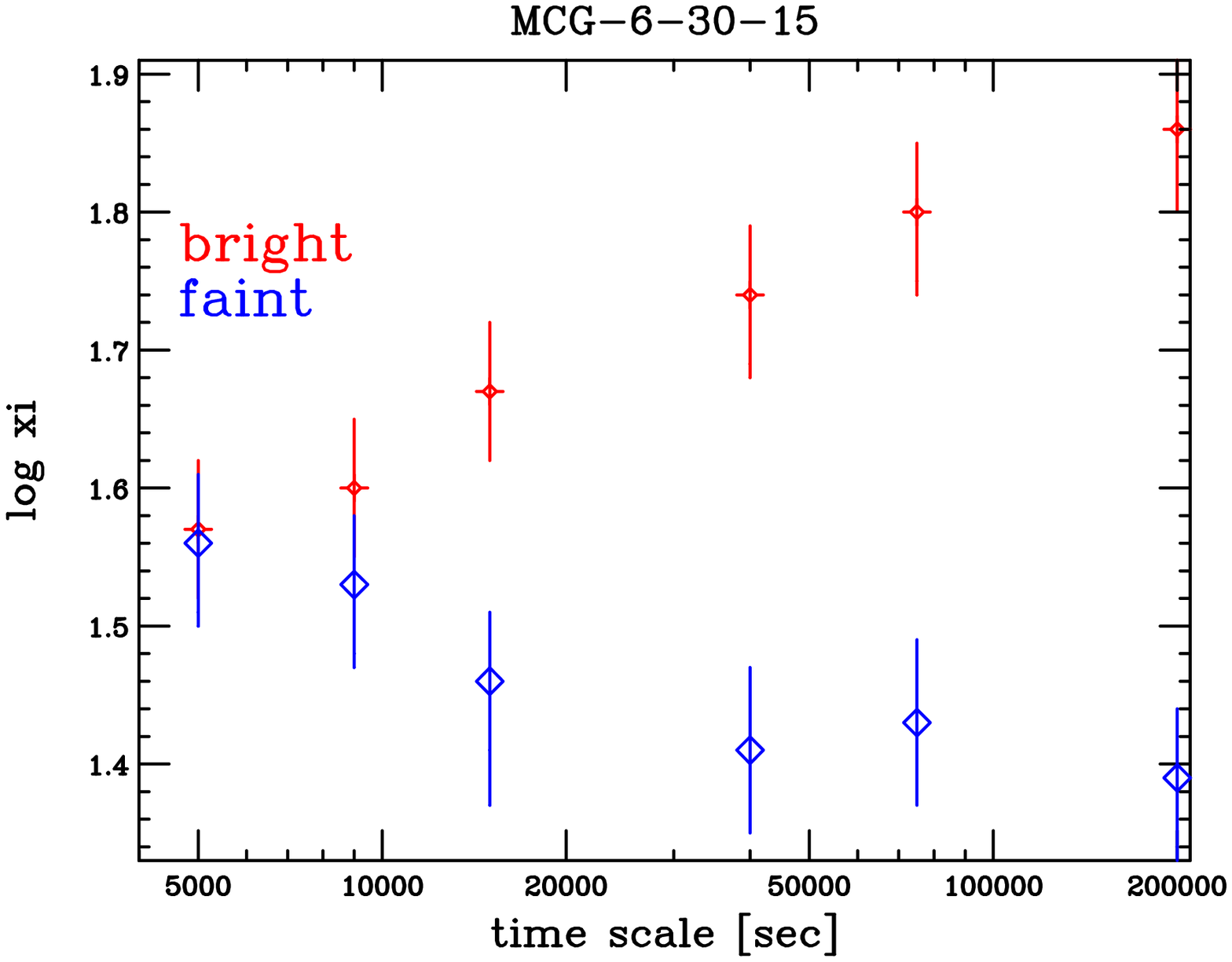}
\FigureFile(80mm,80mm){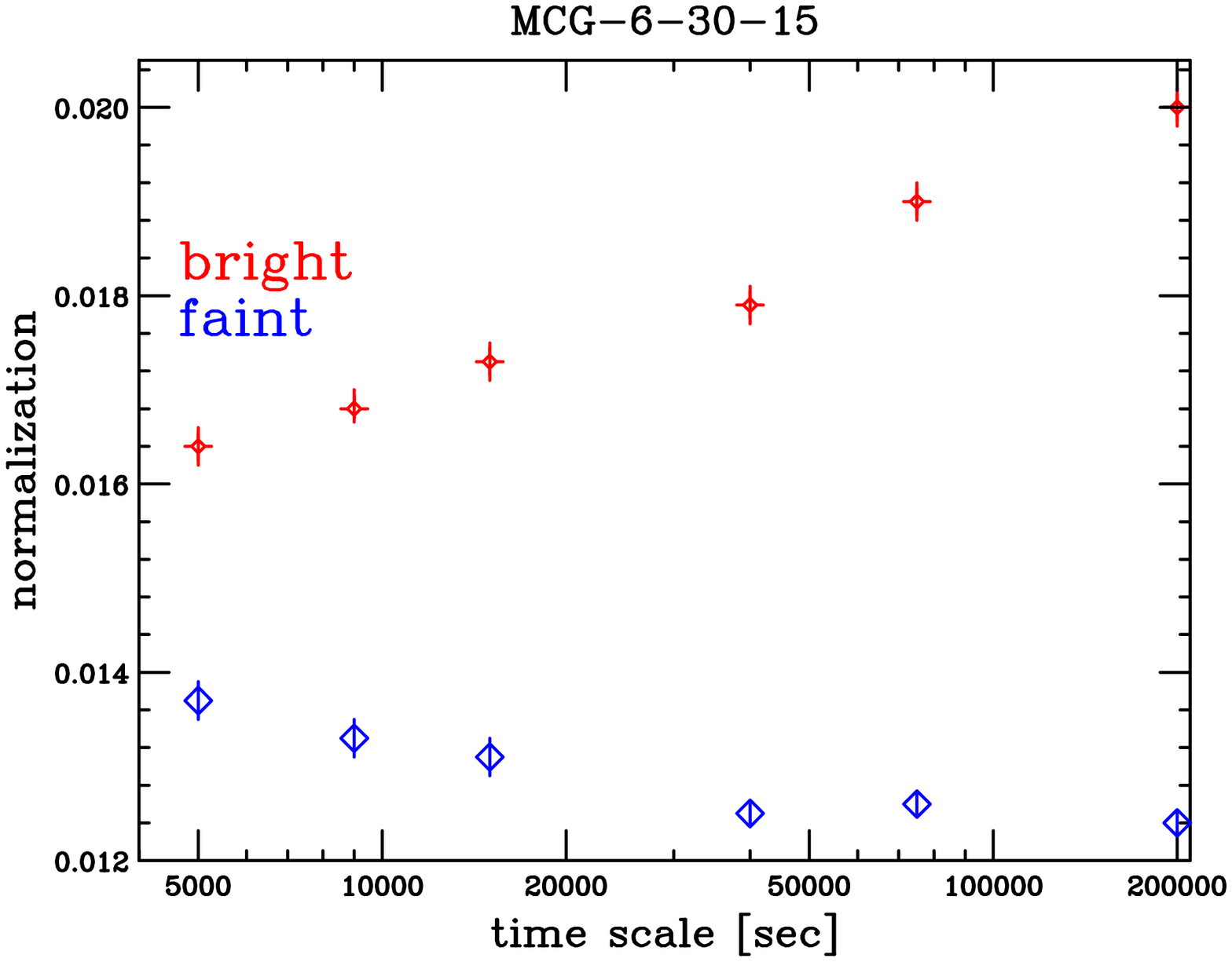}
\FigureFile(80mm,80mm){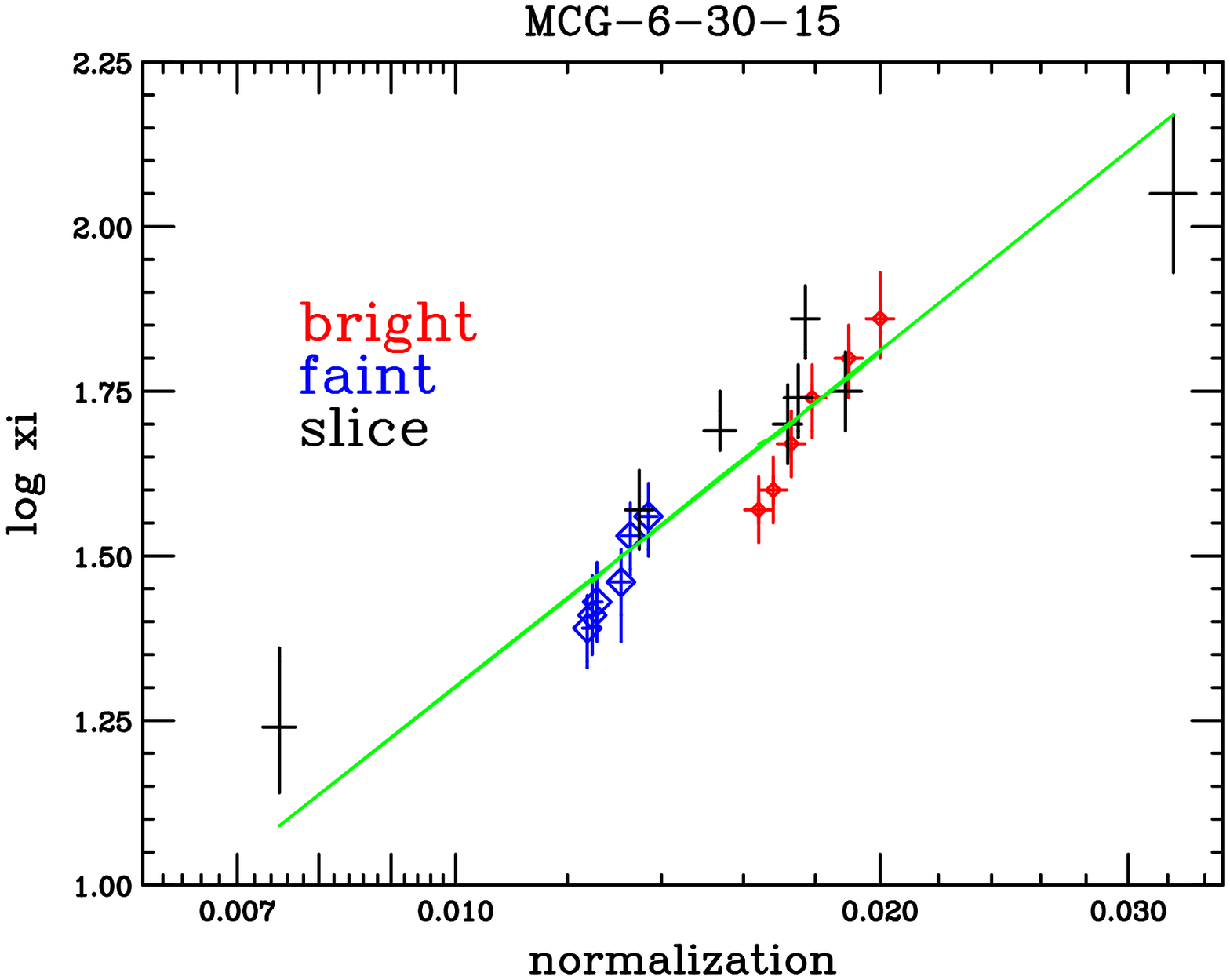}
\end{center}
\caption{Top: Relation between the variation time scale and ionization
 degree for the bright and faint spectra. 
Middle: Relation between the variation time scale and power-law
 normalization for the bright and faint spectra. 
 Bottom: Relation between the ionization degree and
 power-law normalization for the bright, faint and sliced spectra. }
 \label{xi_variation}
\end{figure}

\begin{figure}[htbp]
\begin{center}
\FigureFile(80mm,80mm){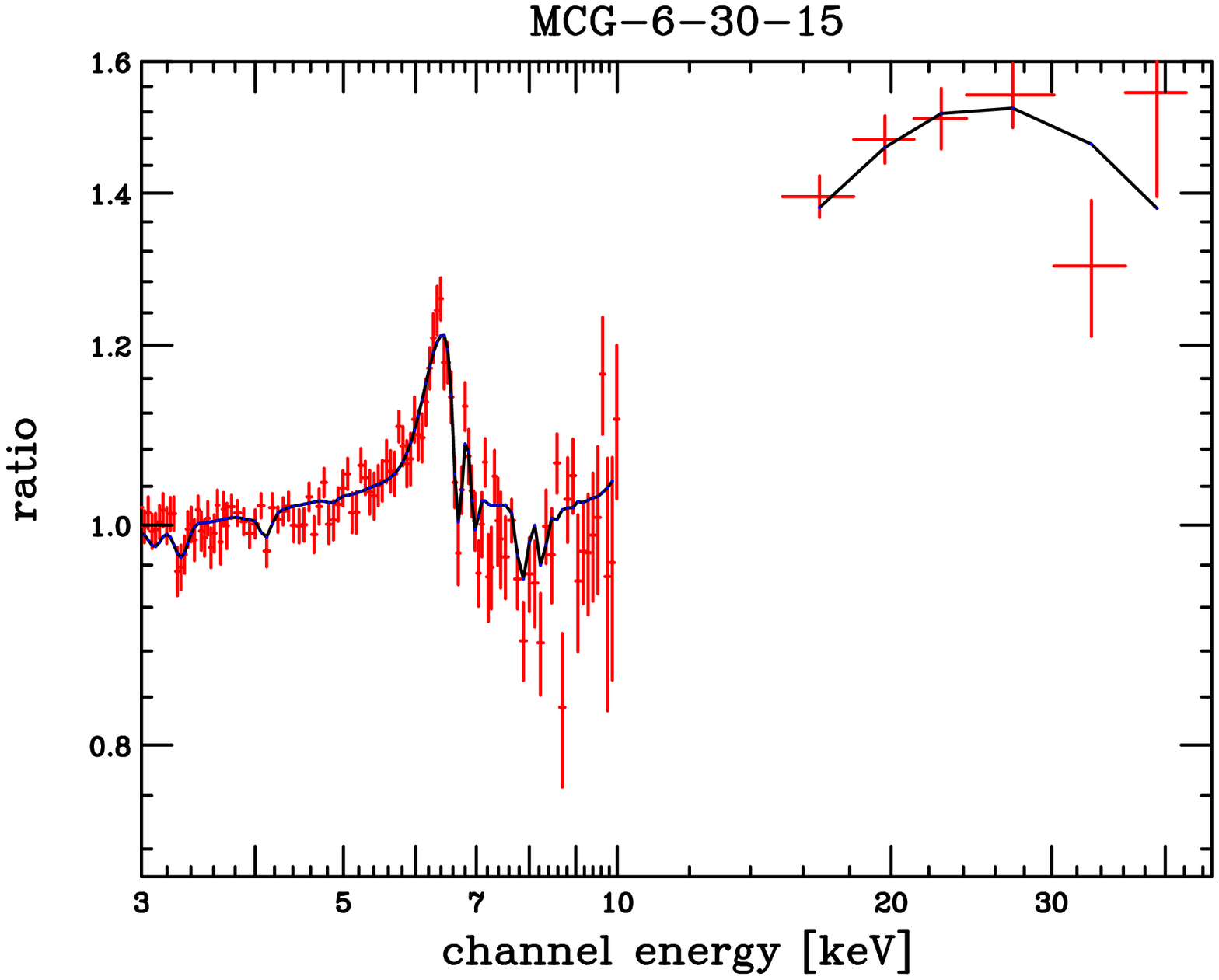}
\end{center}
\caption{Ratio of our best-fit model (black) and the observed data (red) to a 
power-law function.  The power-law slope is determined using the data
only in 3.0--4.0 keV and 7.5--12 keV.  Note that our model includes
a moderately broad iron emission line at 6.42 keV ($1 \sigma=$290 eV), but
not an extremely distorted one with relativistic effects.}
 \label{fig:warm}
\end{figure}

We studied spectral variation of MCG-6-30-15 observed with Suzaku in 
2006 January, and found a clear correlation between  the 6 -- 10 keV
flux and the spectral  ratio of 0.5 -- 3.0 keV/6.0 -- 10 keV 
(Figure \ref{fig:intensityhardness}).  Amplitude of the flux variation and
the accompanying spectral change increases when the variation timescales become
longer from 5 ksec to 200 ksec  (Figure \ref{data345}). 
We fitted the 1 -- 40 keV energy spectra with the spectral model
including a cut-off power-law,  two warm absorbers and neutral reflection
(Figure \ref{fig1}).
The observed spectral variation was almost equally explained by variation of either
photon index of the direct component,  ionization degree 
of the lower-ionized warm absorber, or column density of the warm absorber,
whereas the neutral reflection component was invariable throughout the observation.
We could not clearly judge if the 
intrinsic photon index or  warm absorber parameters are variable,
but it is most physically  reasonable to assume that the photoionization degree increases with increasing
intensities.
We also fit the eight intensity-sliced spectra with the model
varying the power-law normalization and the ionization parameter
(Table \ref{table1_slice}).
In  Fig.$\ref{data12}$, we show the incident
spectral changes for the intensity sliced spectra calculated from the best-fit parameters.

\setlength{\tabcolsep}{2.4pt}
\begin{table*}
\begin{center}
\rotatebox{90}{
\begin{minipage}[t]{\textwidth}
\scriptsize
\caption{Results of spectral fitting in 1--40 keV for the intensity sliced spectra when only the
 power-law  normalization and the ionization degree of the lower-ionized absorber
 are varied. \footnotemark[$*$] }
\begin{tabular}{ccccccccc}
\hline
state  & slice 1 & slice 2 & slice 3 & slice 4 & slice 5 & slice 6 &
 slice 7 & slice 8  \\
\hline
$N_H$ (${10}^{21}$ ${cm}^{-2}$)  &  
		 \multicolumn{8}{c}{2.3$\pm$0.1} \\
\hline
$N_H$  (${10}^{22}$ ${cm}^{-2}$)  &
		 \multicolumn{8}{c}{3.3$\pm$0.9} \\
log $\xi$  & \multicolumn{8}{c}{3.27$^{+0.04}_{-0.06}$} \\
\hline
$N_H$  (${10}^{21}$ ${cm}^{-2}$)  & 
		 \multicolumn{8}{c}{4.1$\pm$0.3} \\
log $\xi$  &  1.24$^{+0.12}_{-0.10}$  &
			 1.57$\pm$0.06 &  1.69$^{+0.06}_{-0.03}$ &
					 1.70$\pm$0.06  &  1.74$^{+0.05}_{-0.06}$ &
							 1.86$^{+0.05}_{-0.06}$ & 1.75$\pm$0.06  & 2.05$\pm$0.12  \\
\hline
Line E (keV) & \multicolumn{8}{c}{6.35 (fix)} \\
sigma (keV) &  \multicolumn{8}{c}{0.01 (fix)} \\
norm (${10}^{-5}$) & \multicolumn{8}{c}{1.7$\pm$0.2} \\
EW (eV) & 68$\pm$6 & 44$\pm$4  &  40$\pm$4  &  36$\pm$3  &  36$\pm$3  &
	35$\pm$3	 &   33$\pm$3   & 21$\pm$2 \\
\hline
cutoffpl K (10$^{-2}$) &  0.75$\pm$0.02 &  1.35$\pm$0.03  &
			 1.54$\pm$0.04  & 1.72$\pm$0.04  &
					 1.75$\pm$0.04	 &  1.77$\pm$0.04 &
							 1.89$\pm$0.05    &  3.23$\pm$0.12  \\
photon index & 
		 \multicolumn{8}{c}{2.04$\pm$0.02} \\
\hline
$E_{cut}$ (keV) & \multicolumn{8}{c}{160 (fix)} \\
cosIncl & \multicolumn{8}{c}{0.866 (fix)} \\
pexrav K (10$^{-2}$)  & 
		 \multicolumn{8}{c}{2.8$^{+0.5}_{-0.3}$} \\
\hline
Line E (keV) &  \multicolumn{8}{c}{2.35$\pm$0.01} \\
sigma (keV) & \multicolumn{8}{c}{0.01 (fix)} \\
norm (${10}^{-5}$) & \multicolumn{8}{c}{-2.4$\pm$0.3} \\
\hline
reduced chisq (d.o.f) & \multicolumn{8}{c}{1.24 (1143)} \\
\hline
\multicolumn{8}{@{}l@{}} {\hbox to 0pt {\parbox{140mm}{\footnotesize
\footnotemark[$*$] 
See the footnote  of Table \protect\ref{table0}.
}\hss}}
\label{table1_slice}
\end{tabular}
\end{minipage}}
\end{center}
\end{table*}

In the top panel of Fig.$\ref{xi_variation}$, 
we show relation between the variation timescales
and ionization degrees of the low-ionized warm absorber.
Also, in the middle panel of Fig.$\ref{xi_variation}$, 
we show the relations between the variation time scales
and the power-law normalizations.
It is obvious that both the ionization degree and
the power-law normalization for the bright state
increases with the variation time-scale, while those for the faint state
decreases  with the variation time-scale.
In the bottom panel of Fig. $\ref{xi_variation}$, we show relation between the power-law normalization
and the ionization degree of the lower-ionized warm absorber,
which indicates a clear correlation as $\log \xi $= (1.7$\pm$0.2) $\times$
log K + (4.7$\pm$0.3). 

Spectral variation in 0.6 -- 1.6 keV can be also explained by
change of the O{\footnotesize VIII} edge depth due to photoionization 
(Figure \ref{below1keVvariation}).
Therefore, we suggest that spectral variation of MCG-6-30-15 in 0.6 keV to 
40 keV on timescales of 5 ksec to 200 ksec is primarily explained by change 
of the ionization degree of  warm absorbers due to photoionization, 
while power-law photon-index and the disk reflection component are invariable.

Note that our model does not include an extremely  broad iron emission line. If the intrinsic
line width is allowed to be free, the central line energy is 6.42$\pm0.06$keV, and the
intrinsic line width is 1$\sigma = 290\pm60$ eV (Figure \ref{fig1-2}).  With the combination
of the cold reflection, mildly broad emission line and warm absorber,
our model can successfully fit the seemingly broad iron line feature
(Figure \ref{fig1-2}).
In Figure \ref{fig:warm}, we show a ratio of our best-fit model
 to a power-law function, as well as a ratio of the observed spectrum
to the same power-law function.
The broad-line like structure down to $\sim$ 4 keV is recognized both
in the model and data, but this is naturally explained by  combination of the disk reflection,
warm absorber and a mildly broad gaussian line that is not significantly red-shifted.
We conclude  that an extremely broad emission line is not required if
multiple warm absorbers are taken into account, which 
agrees with  Miller et al. (2008), while our
model is simpler.


Also we point out that  normalization of the reflection component
requires  $\Omega/2\pi \sim 1$ in our model, 
not $>3$ as Miniutti et al. (2007)  advocates.
Difference is mainly due to the fact that we included warm absorbers, and used
neutral refection without relativistic smearing.
So that the constancy of the reflection component is explained by the
light-bending model, all the disk reflection {\em must}\/  take place
within 3 $r_g$, in which case the reflection normalization requires
$\Omega/2\pi \sim 3$ (Miniutti \& Fabian 2004).
On the other hand, our result favors a smaller solid angle of the reflector. 
Also, invariability of the neutral reflection component suggests that  the reflection takes place 
 far enough from the black hole, so that the intrinsic variation is smeared. 
If that is true, we would expect  a narrow line of the equivalent width $ \sim 150$ eV 
(e.g., George and Fabian 1991) for $\Omega/2\pi \sim 1$,
much larger than the observed value ($\sim$18 eV; Young et al.\ 2005).
If we allow the line width free, we obtain  the equivalent width $\sim$ 110 eV
(table \ref{table0}), which may be reconciled with the reflector having $\Omega/2\pi \sim$ 1.
In this case, we require a mechanism to mildly broaden the line width up to
$1 \sigma \approx 290$ eV.  
If this intrinsic line width is from Keplarian motion,
with $v/c \sim 0.29/6.35$, $v$ is $\sim 14,000$ km/s, which does not seem too
infeasible to be material on the innermost edge of the broad line region.

In summary, we found a characteristic spectral variation of MCG-6-30-15,
that is explained by variation of 
ionization degree of the warm absorber accompanying the flux variation, 
while the neutral reflection
component was invariable throughout the observation. 
Considering the warm absorber, 
the observed energy spectrum is explained by a mildly
broadened iron line, which is not significantly red-shifted, and a neutral 
and constant
disk reflection component of which solid angle is $\Omega/2\pi \sim 1$.
Still, there needs a mechanism to keep the reflection component 
constant while direct component is variable.

\section{Acknowledgement}

The authors would like to thank the referee, Prof.\ Chris Done, for her
valuable suggestions and comments.
This research has made use of public data obtained through the Data ARchives
and Transmission System (DARTS), provided by the  
Institute of Space and Astronautical Science (ISAS), Japan
Aerospace Exploration Agency (JAXA).
For data reduction, we used the software provided by the
High Energy Astrophysics
Science Archive Research Center (HEASARC) at 
NASA/Goddard Space Flight Center. 
T.M. acknowledges Japan Society for the Promotion of Science (JSPS) for 
financial supports via the Research Fellowship for Young Scientists.

\appendix\section{Note Added in the Proof}

After the paper was submitted, the referee pointed out  availability 
of the XSTAR 2.1ln11, and requested to mention  effects of the 
XSTAR version differences.
XSTAR 2.1kn8, which is used throughout the paper,
 does not have all the new K edge structures of Kallman et al.\ 
(2004). These are only incorporated in versions 2.1ln and above.
This makes little difference at high ionisation, but does put
considerably more absorption structures in the K-edge at low
absorptions.

We report a result using XSTAR 2.1ln11, where we fitted the
time-averaged spectra 1--40 keV with a broad emission line 
(${\chi}^2$/d.o.f = 277.2/222). 
We got parameters which agree with those in Table \ref{table0} (left) within
90 \% errors besides the following parameters: $N_H$ of the neutral absorber
is $(1.4\pm0.3) \times10^{21}$cm$^{-2}$ that of higher ionized warm absorber is 
$(23\pm13) \times10^{22}$cm$^{-2}$, and  
that of lower-ionized warm absorber is 
$(4.1\pm1.1) \times10^{21}$cm$^{-2}$. 

In summary, conclusion of our paper is unchanged due to recent version change of  XSTAR.

\end{document}